\documentclass[conference]{IEEEtran}

\usepackage[binary-units]{siunitx}
\usepackage{mathtools}
\usepackage{booktabs}
\usepackage{listings}
\usepackage[hidelinks]{hyperref}

\usepackage{tabularx}
\newcolumntype{Y}{>{\centering\arraybackslash}X}%

\usepackage{csquotes}

\begin{document}

\title{Issues with rounding in the GCC implementation of the ISO 18037:2008 standard fixed-point arithmetic}

\author{
\IEEEauthorblockN{
Mantas Mikaitis}
\IEEEauthorblockA{
Department of Mathematics, The University of Manchester, Manchester, UK \\
Email: mantas.mikaitis@manchester.ac.uk}
}

\maketitle

\begin{abstract}
	We describe various issues caused by the lack of round-to-nearest mode in the \textit{gcc} compiler implementation of the fixed-point arithmetic data types and operations.
	We demonstrate that round-to-nearest is not performed in the conversion of constants, conversion from one numerical type to a less precise type and results of multiplications.
	Furthermore, we show that mixed-precision operations in fixed-point arithmetic lose precision on arguments, even before carrying out arithmetic operations.
	The ISO 18037:2008 standard was created to standardize C language extensions, including fixed-point arithmetic, for embedded systems.
	Embedded systems are usually based on ARM processors, of which approximately 100 billion have been manufactured by now.
	Therefore, the observations about numerical issues that we discuss in this paper can be rather dangerous and are important to address, given the wide ranging type of applications that these embedded systems are running.
\end{abstract}


\IEEEdisplaynontitleabstractindextext

\section{Introduction}

The ISO 18037:2008 standard \cite{iso18037} defines C programming language extensions to support various unconventional features of \textit{embedded} processors.
Embedded processors are usually low power/performance processors found in trains, planes, fabrication equipment and communication devices \cite{marw06}.
Another notable example are battery-powered medical devices using integer processors such as the ARM Cortex-M3 \cite{sdll11}.
One of the main features that the ISO/IEC TR 18037:2008 standard addresses is fixed-point arithmetic and numerical data types for embedded processors.
The standard aims to move away from embedded software designed in assembly languages to a more portable and reusable C programming language, since code is getting bigger and new platforms are rapidly being developed with each new one requiring assembly level changes.

Since processors for embedded systems need to be extremely low power, floating-point hardware support is not affordable and either hardware fixed-point support is provided, or, more commonly, integer arithmetic instructions are used to simulate fixed-point arithmetic.
However, as the standard states, the C programming language does not provide support for any fixed-point arithmetic types which leads to the common solution of handcrafted arithmetic libraries in assembly languages.
The standard aims to improve this situation by defining numerical types and operations that C compilers can support.

In this paper we describe some issues that arise in the \textit{gcc} compiler implementation of fixed-point arithmetic defined by this standard.
Section~\ref{sec:fixed-point-arithmetic}~and~\ref{sec:standard} provides background on fixed-point arithmetic.
Section~\ref{sec:constants} describes the issues with rounding decimal constants to fixed-point data types.
Section~\ref{sec:conversion} describes rounding in conversions between different types.
In Section~\ref{sec:mixed-format} we address mixed-format operations and issues with bit truncation of the arguments, due to limited support for mixed-format operations by \textit{gcc}.
Finally Section~\ref{sec:multiplication} shows that \textit{gcc} does not support round-to-nearest mode on the results of fixed-point multiplication and that the pragma that should enable this rounding mode, as indicated by the standard, does not work.

All of the experiments were compiled with the \textit{gcc} compiler version 9.2.1, using the optimization flag \texttt{-O2} and run on an ARM968 processor.

\section{Fixed-point arithmetic}
\label{sec:fixed-point-arithmetic}

The standard defines multiple numerical types for fixed-point arithmetic in the form $\{s,u\}X.Y$, where $\{s,u\}$ defines whether it is a signed or unsigned format (if signed, 2's complement representation is used), $X$ defines the number of integer bits and $Y$ defines the number of fractional bits.
Machine epsilon of a fixed-point type is defined as $\varepsilon_{\{s,u\}X.Y}=2^{-Y}$, which is the gap between any two neighbouring fixed-point values and is absolute across the dynamic range.
Some notable fixed-point numerical formats supported by \textit{gcc} are: s16.15, u16.16, s0.31, u0.32, s8.7, u8.8, s0.15, u0.16.
The s16.15 representation of the real number $1.5$ has the 14th and 15th bits set to 1 and the others set to 0,  and can be represented as the hexadecimal number 0xC000 or as the integer $2^{15}+2^{14}=49152$.
This can be converted to a decimal value by multiplying it with $\varepsilon_{s16.15}=2^{-15}$, $49152\times2^{-15}=1.5$.
Table~\ref{table:numerical-types-fixed-point} shows examples of some decimal values of the three main numerical types explored in this paper.

\begin{table}[h!]
	\centering
	\caption{Minimum and maximum positive numbers of various 32-bit fixed-point numerical types.}
	\begin{tabularx}{\columnwidth}{rlll} 
		\toprule
		Property & s16.15 & u0.32 & s0.31 \\ \midrule
		Accuracy (abs.)  & $2^{-15}$ & $2^{-32}$ & $2^{-31}$ \\
		Min (exact) & $2^{-15}$ & $2^{-32}$ & $2^{-31}$ \\ 
		Min (approx.) & $0.0000305$ & $2.32 \times 10^{-10}$ & $4.65 \times 10^{-10}$ \\ 
		Max (exact) & $2^{16}-2^{-15}$ & $1-2^{-32}$ & $1-2^{-31}$ \\ 
		Max (approx.) & $65535.999969$ & $0.99...$ & $0.99...$ \\ 
		\bottomrule
	\end{tabularx}
	\label{table:numerical-types-fixed-point}
\end{table}

In terms of rounding, fixed-point arithmetic values can be rounded using the same rounding modes as
floating-point arithmetic, which is defined by the IEEE 754 standard \cite{ieee19}; these are: round-down,
round-up, round-toward-zero and round-to-nearest.
The only difference worth noting is that bit truncation in fixed-point arithmetic is equivalent to round-down mode because of 2's complement notation to represent negative numbers, whereas in floating point it is equivalent to round-toward-zero.

Various libraries have been developed to support fixed-point arithmetic, some of which are based
on the ISO 18037 standard.
\begin{itemize}
\item The \textit{gcc} compiler supports fixed-point arithmetic of the ISO 18037 standard \cite{gcc19c}.
\item The MPLAB XC32 C/C++ compiler, a port of \textit{gcc} for compiling code for the devices developed by a company Microchip \cite{micr16} (unfortunately, support for rounding is not specified).
\item The library called \textit{libfixmath} \cite{goog20} implements a 32-bit fixed-point type with the possibility to control rounding on operations, including bit truncation and round-to-nearest.
\item MATLAB's \textit{Fixed-Point Designer} tool supports fixed-point types with configurable integer and fraction bit-lengths and includes various rounding modes \cite{matl20}.
\end{itemize}
 
\section{ISO 18037:2008 standard}
\label{sec:standard}

The following quotes can be found in Section~4 and Annex A of the ISO standard \cite{iso18037}, dealing with fixed-point number rounding:

\begin{displayquote}
\textbf{Quote 1}: \textit{Conversion of a real numeric value to a fixed-point type may require rounding and/or may overflow.
If the source value cannot be represented exactly by the fixed-point type, the source value is rounded to either the closest fixed-point value greater than the source value (rounded up) or to the closest fixed-point value less than the source value (rounded down).}
\end{displayquote}
Note that Quote~1 can be interpreted to state that one way rounding is suitable, either round-up or round-down, since it does not mention that the decision has to be done based on the bits tha are rounded off.

\begin{displayquote}
\textbf{Quote 2}: \textit{Processors that support fixed-point arithmetic in hardware have no problems in attaining the required precision without loss of speed; however, simulations using integer arithmetic may require for multiplication and division extra instructions to get the correct result; often these additional instructions are not needed if the required precision is 2 ulps. The {\normalfont FX\_FULL\_PRECISION} pragma provides a means to inform the implementation when a program requires full precision for these operations (the state of the {\normalfont FX\_FULL\_PRECISION} pragma is ''on''), or when the relaxed requirements are allowed (the state of the {\normalfont FX\_FULL\_PRECISION} pragma is ''off''). For more discussion on this topic see A.4.
Whether rounding is up or down is implementation-defined and may differ for different values and different situations; an implementation may specify that the rounding is indeterminable.}
\end{displayquote}
Quote~2 talks about a pragma that can be set in order to improve the accuracy of arithmetic operations and
mentions 2 ulp accuracy.
However,  the standard does not mention the error bounds for different rounding modes that can be implemented.
Lastly, this quote has some indication about rounding, and the implication that it may differ for different
values seems to suggest round-to-nearest, but it does not explicitly state this rounding mode and its
maximum error bound of 0.5 ulp that could be achieved if the macro is set.

\begin{displayquote}
\textbf{Quote 3}: \textit{Generally it is required that if a value cannot be represented exactly by the fixed-point type, it should be rounded up or down to the nearest representable value in either direction. It was chosen not to specify this further as there is no common path chosen for this in hardware implementations, so it was decided to leave this implementation defined.}
\end{displayquote}
Quote~3 seems to indicate that rounding should be to one of the two directions, rather than any direction which will give the nearest value.

\begin{displayquote}
\textbf{Quote 4}: \textit{All conversions between a fixed-point type and another arithmetic type (which can be another fixed- point type) are defined. Rounding and overflow are handled according to the usual rules for the destination type. Conversions from a fixed-point to an integer type round toward zero. The rounding of conversions from a fixed-point type to a floating-point type is unspecified.}
\end{displayquote}
This quote provides details about the conversion between different numerical types.
It instructs to apply the usual rules when converting between fixed-point types and any other numerical types,
probably referring to Quote 3.

\begin{displayquote}
\textbf{Quote 5}: \textit{If the result type of an arithmetic operation is a fixed-point type, for operators other than * and /, the calculated result is the mathematically exact result with overflow handling and rounding performed to the full precision of the result type [...]. The * and / operators may return either this rounded result or, depending of the state of the {\normalfont FX\_FULL\_PRECISION} pragma, the closest larger or closest smaller value representable by the result fixed-point type. (Between rounding and this optional adjustment, the multiplication and division operations permit a mathematical error of almost 2 units in the last place of the result type.)}
\end{displayquote}
This quote is very unclear, since it states that the result of addition/subtraction is mathematically exact, but rounded.

\section{Rounding of constants}
\label{sec:constants}

Firstly, we address rounding of constants.
This was commented on previously by us in \cite{hmls20} and also noticed in \cite{tgbd18}.
A constant, for example 0.04, cannot be represented exactly in a finite-precision arithmetic (Table~\ref{table:0.04-constant}) and has to be rounded to the nearest value in the numerical data type of the constant.
For example, the two nearest values in the integer representation in s16.15 are $\lfloor\frac{0.04}{2^{-15}}\rfloor=1310$ and $\lceil\frac{0.04}{2^{-15}}\rceil=1311$, produced by round-down and round-up respectively.
These correspond to the real values of $0.039978...$ and $0.040008...$.
However, since $\frac{0.04}{2^{-15}}=1310.72$, it makes most sense to represent $0.04$ as $\lceil\frac{0.04}{2^{-15}}\rceil=1311$, since it is closer to the real value of $0.04$.
That is, round $0.04$ to the nearest s16.15 value (or any other given fixed-point format that is being used to store the constant).
This operation is done on compilation, when the constant is written into the memory by the compiler, and therefore no run-time performance penalty is incurred.
Unfortunately, we found that this was not done by the \textit{gcc} compiler, which resulted in large total errors due to magnification of these small errors in the constants, for example in Ordinary Differential Equations (ODE) solvers run using fixed-point arithmetic with the \textit{gcc} compiler \cite{tgbd18, hofu15}.
The code for this is
\begin{lstlisting}
   accum a = 0.04k;
\end{lstlisting}
where the letter \texttt{k} is used to indicate that this constant is in s16.15 format (not necessary to use in this context since the destination format is known but we chose to use it for demonstration).
The \textbf{accum} data type is another name in C for the s16.15 data type.

We believe this to be an issue due to Quote~2 --- the pragma that is defined there should only be applied to control run-time performance, that is, rounding of various values that come up at run time, not at compile time.
On compilation we expect all the constants to be rounded to the nearest representable values irrespective of the run-time accuracy settings.
And in general, we found that the pragma FX\_FULL\_PRECISION does not have any effect in \textit{gcc} and does not turn on rounding neither on compilation nor run time.

\begin{table}[h!]
	\centering
	\caption{Values of a constant 0.04 in different data types}
	\begin{tabularx}{\columnwidth}{rXX} 
		\toprule
		Data type & round-to-nearest & next nearest \\ \midrule
		s16.15  &  $0.040008544921875$ & $0.03997802734375$ \\ 
        s0.31 & $0.04000000003725...$ & $0.0399999995715...$ \\
		u0.32 & $0.04000000003725...$ & $0.0399999998044...$ \\
		binary32 &  $0.03999999910593...$ & $0.0400000028312...$ \\
		\bottomrule
	\end{tabularx}
	\label{table:0.04-constant}
\end{table}

\section{Rounding on conversion}
\label{sec:conversion}

Here we show that round-to-nearest is not applied when converting to a fixed-point type a numerical value that is held in a more precise data type.
First, we try to convert a value held in s0.31 to s16.15.
We choose a value that is smaller than the smallest value representable in s16.15: $2^{-16}+2^{-17}=0.00002288818359375=0.75\varepsilon_{s16.15}$, where $\varepsilon_{s16.15}=2^{-15}$.
In C code we write:

\begin{lstlisting}
  long fract a = 2.288818359375E-5lr;
  accum b = a;
\end{lstlisting}
Here \textbf{long fract} is another name for s0.31 and \textbf{accum} for s16.15.
The letters \texttt{lr} next to the constant tell the compiler that this is a s0.31 constant, as defined in the ISO standard.
Once this code is executed, \textit{b} evaluates to 0, rather than the nearest representable value of $2^{-15}$, therefore round-to-nearest is not performed on conversion --- round-down, or bit truncation, is performed instead.

Another test that demonstrates this involves conversion from the single precision floating-point format binary32, defined in the IEEE 754 standard \cite{ieee19}, to s16.15.
\begin{lstlisting}
  float a = 0.04;
  accum b = a;
\end{lstlisting}
This uses the same constant that we have used in the Section~\ref{sec:constants}, which is not rounded to the nearest fixed-point value when specified as a decimal value 0.04 in the source code as was shown in Section~\ref{sec:constants}.
In this case binary32 approximation of 0.04 is more accurate than the s16.15 approximation, so the value of 0.04 held as binary32 should be rounded to the nearest value in s16.15. However, $b$ still evaluates to the value below 0.04, meaning that upon conversion from binary32 to s16.15 round-down is used instead of round-to-nearest.

Therefore, conversion of fixed-point values does not implement round-to-nearest to minimize the conversion
error and either follows the vague specification of the standard in Quotes~1,~3,~and~4 or simply performs
bit truncation by shifting.
Our recommendation is that the specification in Quote~2 should be implemented, with the pragma enabling
round-to-nearest on conversions.

\section{Rounding of arguments in mixed-format operations}
\label{sec:mixed-format}

We have observed multiple fixed-point arithmetic routines in the assembly from \textit{gcc} with some loss of
precision and speed, for example the multiplication of a value in s16.15 by u0.32 is performed as the multiplication
of two s32.31 values, or the multiplication of s16.15 by u0.16 is performed as the multiplication of two s16.15 values.
This is achieved by converting all the arguments to the common internal format, which means that u0.32 argument in
the former case is converted to s32.31, and u0.16 argument in the latter case to s16.15 (one bit less precision in the
fractional part).
This causes loss of precision on conversion in the arguments, even before multiplication is performed, and the main
reason is that \textit{gcc} does not support mixed-format multipliers directly, as indicated by a list of internal compiler functions for performing fixed-point arithmetic operations \cite{gcc19a}.
A test for this is as follows:
\begin{lstlisting}
  unsigned long fract a = pow(2,-32);
  accum b = 65535k;
  unsigned long fract c = a * b;
\end{lstlisting}
We chose $a=\varepsilon_{u0.32}=2^{-32}$ since that is the smallest value representable by u0.32 (only the least significant bit set) and $b=65535$ the largest integer value representable by s16.15.
In this scenario we expect to get $c = 65535 \times a$, however we get $c=0$ because the last bit of $a$ is dropped before the multiplication takes place, causing $a=0$.
Same issue happens irrespective of what $b$ is set to.
Furthermore, we can enclose this code in a conditional execution that checks the values of \textit{a} and \textit{b} and it executes the conditional code and incorrectly updates \textit{c} to 0, overwriting it's previous value:
\begin{lstlisting}
  unsigned long fract a = pow(2,-32);
  accum b = 65535k;
  unsigned long fract c = 0.8ulr;
  if (a > 0 && b > 1)
    c = a * b;
\end{lstlisting}
Lastly, if we modify the code as follows:
\begin{lstlisting}
  long fract a = pow(2,-31);
  long fract b = -1lr;
  long fract c = a * b;
\end{lstlisting}
In this case we are using signed fractional type s0.31 so that we can represent $-1$.
Running this testcase we do not get $c=0$ but a correct multiplication result of $c=-2^{-31}$.
Both this and the previous testcase are quite similar --- multiplying a very small value by a value that
is not smaller than $1$ in magnitude, expectation is that this code will scale $a$, returning $|c| \geq |a|$.
However, the first provides unexpected result due to the conversion of $a$ and $b$ to the common numerical type s32.31, as outlined above, and the second works as expected as no conversion is needed.
This leads to a major problem: we know that \textit{a} is not zero, but multiplying it by a non-zero value with a magnitude larger than 1 sometimes can give an answer of 0 and sometimes a correct answer, depending on the numerical types used to store $a$ and $b$.
For most of the users who do not necessarily think about how exactly arithmetic is performed at the lowest level, this behaviour would be and potentially is very puzzling.

\section{Rounding of multiplication results}
\label{sec:multiplication}

In this section we show that there is no support for round-to-nearest in arithmetic operations with fixed-point numbers.
Specifically, we test multiplication, which multiplies the two arguments and returns a value in a more precise fixed-point format.
The result held in extended precision in most cases requires rounding in order to convert to the format of one of the arguments.
A simple test is to declare two s16.15 values, $a=3 \times \varepsilon_{s16.15}=0.000091552734375$ and $b=0.25$.
This should give us $0.25\times3\varepsilon_{s16.15}=\frac{3}{4}\varepsilon_{s16.15}$ which should round to a nearest value of $\varepsilon_{s16.15}$.
The code for this is:
\begin{lstlisting}
  accum a = 0.000091552734375k;
  accum b = 0.25k;
  accum c = a * b;
\end{lstlisting}
In this piece of code $c$ evaluates to 0, which means that the result $\frac{3}{4}\varepsilon_{s16.15}$ is rounded down to 0 rather than the closest value of $\varepsilon_{s16.15}$, most likely as a result of bit truncation
that happens when the product in full precision (which has to be stored in the s32.30 format) is shifted right 15 steps to convert it to s16.15.
The pragma that is described by Quote~2 does not change the rounding mode.

\section{Conclusion}

We have shown various numerical accuracy issues in the \textit{gcc} compiler implementation of the standardized fixed-point arithmetic \cite{iso18037}.
The main issue is lack of rounding in decimal to fixed-point conversion, generally any format to fixed-point conversion and arithmetic operations such as multiplication.
Furthermore, there is precision loss in the arguments in mixed-format arithmetic operations.
In our understanding, these software bugs exist both because of the vague specifications of various fixed-point
properties and required features in the ISO 18037 standard, and lack of support of different features of the
fixed-point arithmetic in \textit{gcc}.
This arithmetic in \textit{gcc} should be carefully reimplemented taking care of various edge cases and all possible mixed-format combinations to support the embedded systems community.

In summary, the current paper should inform the embedded systems community about the numerical accuracy problems in the current implementation of the \textit{gcc} fixed-point arithmetic, as well as help identify and understand numerical problems in their codes.

\section{Acknowledgements}

The author thanks to Massimiliano Fasi and Nicholas J. Higham for their feedback on the early drafts of the manuscript. The author was supported by an EPSRC Doctoral Prize Fellowship.

\bibliographystyle{IEEEtran}
\bibliography{bibliography}

\end{document}